# The 3.6 metre Devasthal Optical Telescope: from inception to realization

*Ram Sagar*, Brijesh Kumar and Amitesh Omar*

*India's largest 3.6 metre Devasthal Optical Telescope (DOT) was commissioned in 2016, though the idea of building it germinated way back in 1976. This article provides research accounts as well as glimpses of its nearly four decades of journey. After a decade of site surveys, Devasthal in the central Himalayan region of Kumaon, Uttarkhand was identified. Thereafter, a detailed site characterization was conducted and project approvals were obtained. The telescope is designed to be a technologically advanced optical astronomy instrument. It has been demonstrated to resolve a binary star having angular separation of 0.4 arc-sec. After technical activation of the telescope on 30 March 2016, it has been in regular use for testing various back-end instruments as well as for optical and near-infrared observations of celestial objects. Back-end instruments used for these observations are 4K × 4K CCD IMAGER, faint object imager-cum-spectrograph and TIFR near-infrared camera-II. A few published science results based on the observations made with the telescope are also presented. Furthermore, routine observations show that for a good fraction of observing time the telescope provides sky images of sub-arc second resolution at optical and near-infrared wavelengths. This indicates that the extreme care taken in the design and construction of the telescope dome building has been rewarding, since the as-built thermal mass contributes minimally so as not to degrade the natural atmospheric seeing measured at Devasthal about two decades ago during 1997–99 using differential image motion monitor. The overall on-site performance of the telescope is found to be excellent and at par with the performance of other similar telescopes located over the globe.*

**Keywords:** History, optical telescope, optical observatory, site characterization, sky performance.

OPTICAL astronomy is as old as human civilization as our ancestors started looking towards the sky with naked eyes, which acted as a telescope of pupil size ~6–7 mm, while the retina worked as the detector and the brain functioned as a computer. However, a major breakthrough in our understanding of the Universe came about 400 years ago in the year 1609, when Galileo Galilee, for the first time, viewed and studied planets of our solar system with a small (~4 cm) optical telescope. Since then optical telescopes have played important roles in deciphering secrets of the Universe. All over the globe, astronomy and astrophysics is generally nurtured by the Government agencies as a branch of pure science. Most of the astrophysical research in India is supported by various ministries of the Government of India (GoI), such as the Department of Science and Technology (DST), Department of Atomic Energy, Department of Space, and Ministry of Human Resource Development, and also by several state government departments.

The first 1 m aperture optical telescope in India was established in 1972 at Kavalur, Tamil Nadu, and thereafter, a few more 1–2 m-class optical telescopes were established in the country, including the 2.3 m Vainu Bappu Telescope at Kavalur. These optical telescopes are primarily used for photometric and low-resolution (~2000) spectroscopic studies of relatively brighter celestial sources ($V \leq 21$ mag). Their global relevance in the era of multiwavelength astronomy has been highlighted elsewhere[1]. The large size optical telescopes help in probing fainter and distant celestial sources. However, installation of such telescopes takes several years as it involves time-consuming activities like identification of site for locating the telescope, finalization of technical parameters of the telescope, and completion of financial and administrative approvals from the competent authorities. The idea of installing a 4 m-class optical telescope in India, as a major national observing facility by involving other interested institutions, germinated in 1976, when the

Ram Sagar, Brijesh Kumar and Amitesh Omar are in the Aryabhatta Research Institute of Observational Sciences, Manora Peak, Nainital 263 001, India; Ram Sagar is also with the Indian Institute of Astrophysics, Sarajapur Road, Koramangala, Bengaluru 560 034, India.
*For correspondence. (e-mail: ramsagar@iiap.res.in)

**Table 1.** Chronological list of events in year(s) related to the 3.6 m DOT project

| Duration | Description of activities |
|---|---|
| 1976 | Uttar Pradesh (UP) State Government provided initial in-principal approval. |
| 1980–1990 | Preliminary site survey was carried out, but the detailed project report (DPR) prepared by Uttar Pradesh State Observatory (UPSO) and Bhabha Atomic Research Center, Mumbai was deferred by the Planning Commission, Government of India (GoI). |
| 1996–2001 | Characterization of Devasthal sky in collaboration with Indian Institute of Astrophysics, Bengaluru and Tata Institute of Fundamental Research (TIFR), Mumbai. |
| 1996–2010 | Land acquisition from UP and Uttarakhand state governments. |
| 1997–2000 | The Planning Commission, GoI approved UPSO and TIFR DPR with equal partnership. |
| 2001 | The project was got deferred by the Uttarakhand State Government. |
| 2004 | Cabinet of GoI approved the project and also formed ARIES under DST. |
| 2005–2006 | Expenditure Finance Committee (EFC), GoI; Indian astronomical community and Governing Council of ARIES approved the project and formed a nine-member Project Management Board; Belgium joined the project by contributing one-time fixed amount of 2 million Euros. |
| 2006–2007 | Telescope specifications were frozen and contract was awarded to Advanced Mechanical Optical System (AMOS). |
| 2007–2008 | AMOS sub-contracted supply of M1 and M2 mirrors to Lytkarino optical glass factory (LZOS); Astro-Sital M1 blank could not meet the specified optical quality and its replacement with Zerodur increased the project cost. |
| 2009-2011 | Revised EFC proposal costing ~120 crores Indian rupees plus 2 million Euros as Belgian contribution was approved by GoI; LZOS supplied M1 and M2 mirrors. |
| 2012–2015 | Factory test at AMOS; transportation of telescope parts, completion of telescope building, installation and successful sky verification according to specifications. |
| 2016 | Telescope technically activated by the Premiers of both India and Belgium on 30 March 2016. |

Uttar Pradesh State Observatory (UPSO) got in-principal approval from the Uttar Pradesh (UP) State Government[2]. However, it was realized recently in 2016 by establishing the 3.6 m Devasthal Optical Telescope (DOT)[3], the largest optical telescope in the country.

This article provides research and historical accounts on how the 3.6 m DOT observing facility was developed and is being used. Historical accounts, development of infrastructure at Devasthal, and various administrative and financial approvals for the project are described. Importance of site survey for installation of an optical telescope and modern equipment used for astronomical characterization of the Devasthal site are also discussed. A brief description of all these activities followed by a technical description of the telescope, back-end instruments, research observations, and a summary and future outlook are also provided.

## Historical account of the 3.6 m DOT project

Table 1 summarizes various approvals and activities related to the telescope project. It took about two decades (1980–2001) for site characterization and then further 15 years for the development, manufacturing, installation, commissioning and sky performance testing of the telescope. The UPSO initiated site survey for locating the telescope in 1980, but the detailed project report (DPR) prepared jointly with Bhabha Atomic Research Center (BARC), Mumbai, was deferred by the planning commission, GoI, and the project activities were practically abandoned in 1990 (ref. 4). After a gap of about 6 years, astronomical characterization of Devasthal site was started in collaboration with the Indian Institute of Astrophysics (IIA), Bengaluru, and the Tata Institute of Fundamental Research (TIFR), Mumbai. A new DPR prepared jointly by UPSO and TIFR was approved by the Planning Commission, GoI, in 1998. Under the Chairmanship of Director, TIFR, a Project Management Board (PMB) consisting of members from both the UP State Government and TIFR was formed. The first meeting of PMB was held in June 2000 at UPSO, Nainital. However, a few months later, on 9 November 2000, formation of Uttarakhand State by geographical bifurcation of UP was announced by GoI, and the Observatory became part of the newly formed state. In January 2001, the Uttarakhand State Government approached DST for taking over the State Observatory because the development of pure science is a mandate of GoI. TIFR was informed about this development in the second PMB meeting held in June 2001. The authorities of TIFR then decided to withdraw from the telescope project. Subsequently, Aryabhatta Research Institute of Observational Sciences (ARIES), an autonomous research institution under DST was formed through a Cabinet decision of GoI taken in its meeting held in January 2004, with an objective of its seamless integration with other similar national institutions for making best use of its scientific potential[5,6]. Since the large-telescope project was already part of the January 2004 Cabinet approval, ARIES took initiative to develop a large-sized optical telescope as a national facility. The Governing Council (GC) of ARIES in September 2004 resolved to obtain post-facto financial approval for the telescope project as a procedural compliance of the January 2004 Cabinet approval. The project was also endorsed by the astronomical community of India during the 23rd Annual Meeting of the Astronomical Society of India (ASI), held in Nainital in February 2005. The key areas of research identified at the ASI meeting included asteroseismology, search for extra-solar planets, study of star formation and stellar magnetism, probing highly energetic sources, and optical study of X-ray and radio sources. One of the main objectives of the optical telescope project was to carry out follow-up studies of sources identified at radio wavelengths by the Giant Metre-wave Radio Telescope (GMRT), and at UV/X-ray wavelengths by AstroSat, the first Indian multi-wavelength space observatory.

Meanwhile, the astronomical community of Belgium also showed interest in joining the project[7,8] by contributing 2 million Euros. Thereafter, a fresh DPR was prepared and presented to the Expenditure and Finance Committee (EFC) of DST, which approved the project in June 2006. The GC of ARIES in its meeting held in October 2006 also approved the telescope project and formed a 9-member PMB comprising Indian experts in the field of astronomical instrumentation. Based on broad telescope specifications defined by the PMB and after the due tendering process, ARIES awarded the contract for design, manufacture, integration, testing, supply and installation of a 3.6 m aperture size modern optical telescope at Devasthal to the Advanced Mechanical and Optical System (AMOS), Belgium in March 2007. AMOS sub-contracted supply of primary (M1) and secondary (M2) mirror blanks as well as their surface polishing to the Lytkarino Optical Glass Factory (LZOS), Russia. Unfortunately, the initially planned LZOS Astro-Sitall M1 blank faced poorly in the optical quality tests carried out by AMOS in March 2008. Consequently, after approval from the competent authorities, Zerodur M1 blank was procured from Schott, Germany, by the end of 2008 and was sent to LZOS for surface polishing. This increased the project cost in comparison to the cost approval given by EFC in June 2006. Hence, a revised EFC proposal was prepared during 2009–10, and financial approval for the project costing a total of about Indian Rupees 120 crores plus 2 million Euros as Belgian contribution was given by the competent authority in February 2011. This revised cost included the telescope-building development work, infrastructural work at site such as roads and power, and other miscellaneous project expenses.

The LZOS completed surface polishing of both mirrors M1 and M2 of the telescope in 2010, and published as-built optical quality of the mirrors[9]. The AMOS and ARIES carried out assembly, integration and first-light verification tests of the telescope at factory in 2012, and published the results[10–12]. Parts of the telescope were transported by sea and by road from Belgium to India in 2013. Integration of the telescope was carried out during October 2014 to February 2015.

## *Identification of potential site for optical astronomy*

Before installing an optical telescope at a site, astronomical characterizations of site conditions are essential as they provide valuable information about telescope house design and its expected observing efficiency. This can be understood from the fact that deep imaging capacity of a telescope is $\propto(A_{\mathrm{eff}}/\varepsilon_{\mathrm{D}})$, where $A_{\mathrm{eff}}$ is the telescope light-gathering power, including the losses due to optics and the quantum efficiency of the detector used at the focus of telescope, and $\varepsilon_{\mathrm{D}}$ is the solid angle formed by the seeing for a ground-based optical telescope of size ≥15 cm (ref. 13). Consequently, a 2 m telescope located at a site with a seeing of say 0″.5 performs like a 4 m telescope installed at a site having seeing of 1″, if other conditions are identical. For making the best use of the optical telescope, the site should have maximum possible number of usable nights in a year, dark and transparent skies, low perceptible water vapour content, and small changes in the ambient temperature during night. In addition, the site should be far away from human activities so that the contribution from city lights to the night-sky brightness could be minimal. At the same time, one has to consider availability of water and power to the site, as well as its approachability so that infrastructural development and telescope operation do not become too expensive. Some compromise between all these factors is always made while choosing an observatory site. The cost involved in providing easy accessibility to an otherwise excellent site often comes in the way of its choice. Most of the best optical observatories over the globe are located at an altitude of ≥2 km in the subtropical zone within 25°–35° lat. from the equator, either on islands such as Hawaii and La Palma, or in coastal areas such as Chilean and Midwest American. The 2 m Himalayan Chandra Telescope is located at an altitude of 4.5 km in Changthang region, Leh district, Jammu & Kashmir[14].

In light of the above, the site survey work for locating the telescope started during 1980s, mainly in Kumaon and Garhwal regions of central Himalaya. Using Survey of India maps, a total of 36 prospective sites were identified. Then first-hand information such as distance from the road, availability of water and power line, obstruction due to surrounding hills was collected during 1981–82 and five sites, viz. Gananath, Mornaula, Chaukori, Jaurasi and Devasthal, all having altitude ≥2 km, were shortlisted for further investigations. Meteorological data were collected during 1982–1991 using thermograph, hygrograph, barograph, sunshine recorder, rain gauge, snow gauge, wind speed and direction recorder. The cloud coverage was recorded visually. Key results of these observations have been published[15]. Star trails were recorded on a film using a Kodak SLR, 35 mm camera mounted on polar star trail telescopes at Devasthal and Gananath sites. The full width at half maximum (FWHM) of star trail profile works as a proxy for seeing, and it was measured at a number of points using a micro-densitometer. These measurements showed that most of the time FWHM at Devasthal was better than that at Gananath[4]. Devasthal was finally identified as a potential site for setting up a 4 m class-optical telescope facility. Figure 1 shows the geographical location and topographic contour map of the Devasthal site. Its topography indicates that the Devasthal is the highest peak in the region of >10 km range. Figure 1 also shows that Devasthal is at a point with sharp altitude gradient towards the southwest, the prevailing incoming wind direction at the site. Therefore, this location is expected to provide better seeing for astronomical observations.

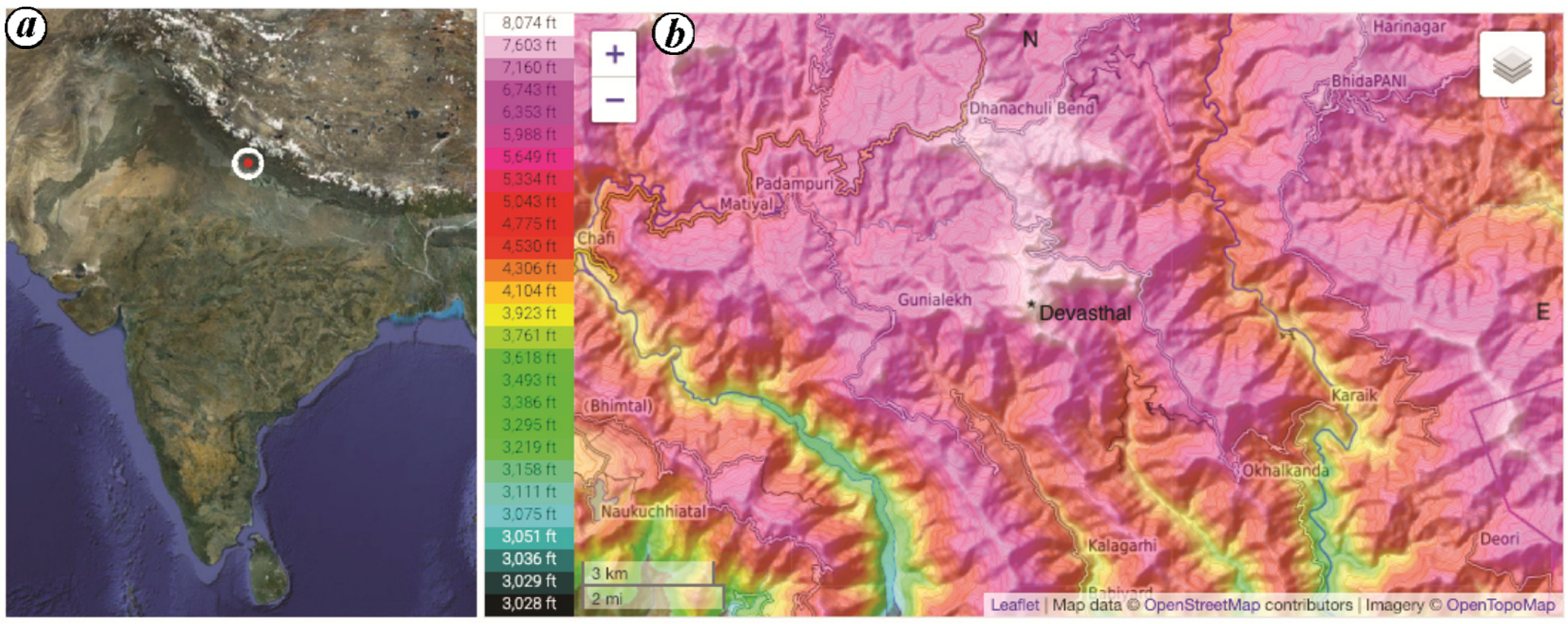

**Figure 1.** ***a***, The location of Devasthal on the Indian sub-continent is marked with a red dot encircled in white colour. ***b***, The topographic contour map showing the location of Devasthal, the highest peak in the region and its immediate surroundings. North is up and East is to the right in this map. The prevailing incoming wind direction is southwesterly.

### *Acquisition of land near mountain top*

In order to build connecting road and infrastructure for the Devasthal observatory, the process of land acquisition started in 1996. The observatory is spread across area of 4.1692 ha, located in South Gola forest region of Nainital district, out of which, 3.9192 ha is in Kullauri Van Panchayat while 0.25 ha is in Surang Van Panchayat. This forestland was transferred by the UP State Government in 1999. All telescope buildings, namely 4 m International Liquid Mirror Telescope (ILMT), 1.3 m Devasthal Fast Optical Telescope (DFOT) and 3.6 m DOT, and support infrastructural facilities like guest rooms, road linking to various facilities, power stations, data centre, etc. are located in this region[16]. Three kilometre long metaled road links the Uttarakhand state government road from Jadapani to the Devasthal Observatory. It is located in the Khutiakhal and Darmane Rata Van Panchayats of North Gola forest region of Nainital district, and covers an area of 0.8734 ha, of which 0.3150 ha was transferred by the UP State Government in 1999, while the remaining area of 0.5584 ha was transferred by the Uttarakhand State Government in 2010.

## Characterization of Devasthal site

The FWHM of a point source image formed at the focal plane of a telescope is affected not only by the Earth atmospheric seeing, but also by the motion of the telescope, which includes wind shaking, guiding, dome effect, etc. For seeing measurements, it is therefore essential to eliminate the effects due to motion of the telescope, and this is achieved in the Differential Image Motion Monitor (DIMM) set-up. The principle of DIMM is to produce twin images of a star with the same telescope through two entrance pupils separated by a fixed distance. It, thus, works on the principle of measuring image motion at the focal plane of a telescope, which eliminates the effects due to motion of the telescope and hence enables us to measure the contributions of Earth's atmosphere to image degradation.

The European Southern Observatory developed a DIMM instrument and used it extensively for site survey in Chile during late 1980s (ref. 17). Following this, 52 cm optical telescope was installed at base-camp and 38 cm optical telescope was installed at hilltop of the Devasthal Observatory[15]. Figure 2 shows a ray diagram of DIMM along with sizes of masks having two circular holes and separation between them. Measurements of relative motions between the two images provide variations in the relative slope of the wavefront at the two circular holes, as the possibility of changes in the wavefront is negligible during the short exposures (10 ms) used in the DIMM observations. Image motion data obtained in this way were analysed on-line to provide seeing measurements. Further details of the instrument and key scientific results obtained with these extended observations are published elsewhere[15,18]. The seeing measurements at the hilltop were carried out during 1998–99 using a 38 cm optical telescope with mirror M1 ~2 m above the ground. The observations carried out over 80 nights yielded a median seeing of 1″.1, a seeing of <1″ for 35% and of <0″.8 for 10% of the time, and a minimum recorded seeing of 0″.38.

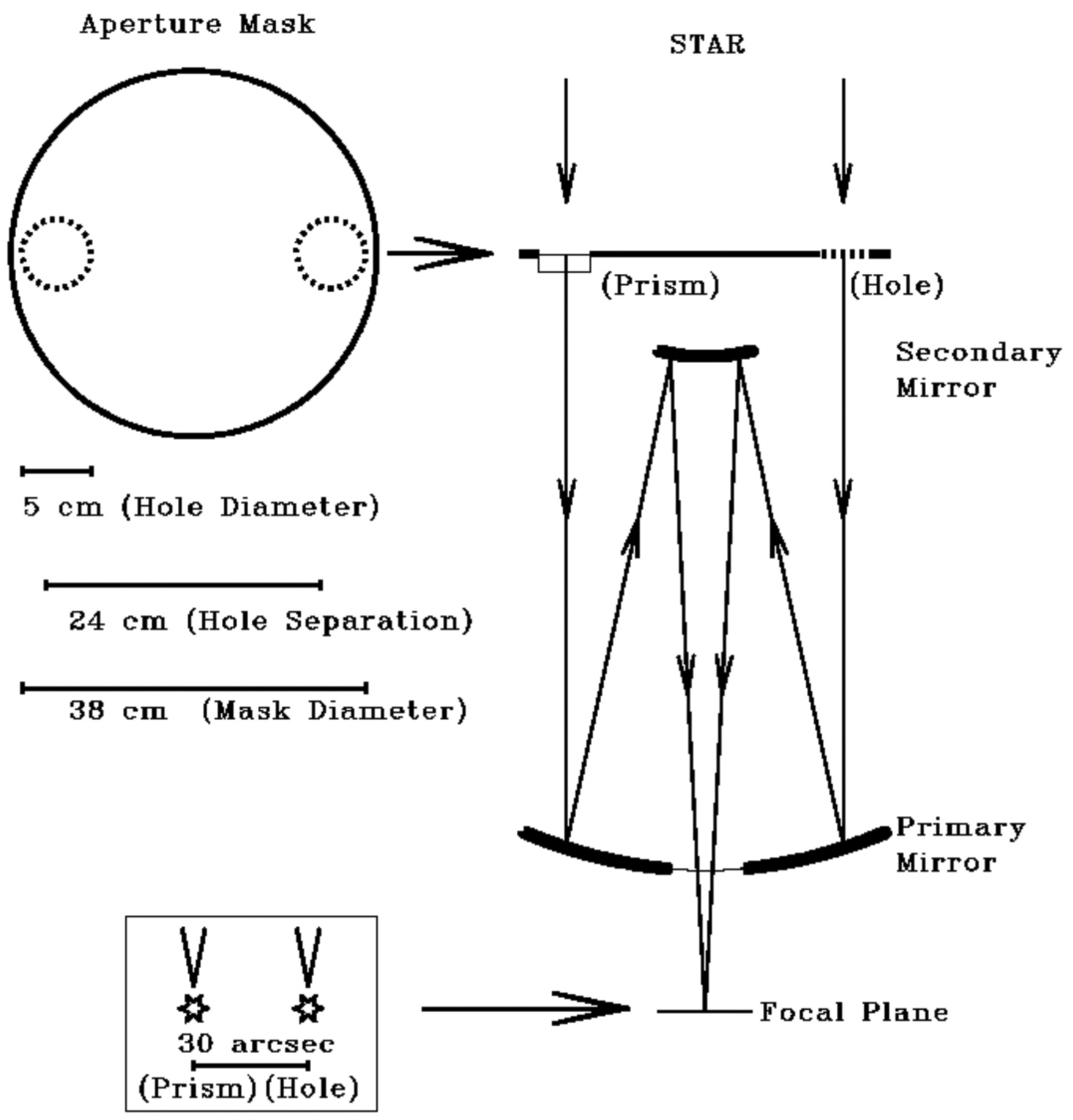

**Figure 2.** A ray diagram of the differential image motion monitor (DIMM) set-up mounted on the 38 cm telescope. The two circular holes of 5 cm diameter in the aperture mask were separated by 24 cm. The same set up was used for 52-cm telescope and the corresponding numbers were 6 cm and 40 cm respectively. The hole mounted with a prism deviates the incoming star light by about 30″ in the direction of the line joining the centres of the two holes. The Santa Barbara Instrument group ST-4 auto-guiding CCD camera, controlled by a PC, was mounted at the focal plane of the telescope for recording the two images of the same star.

The natural seeing at the telescope deteriorates due to temperature gradients produced by the immediate surroundings. In order to estimate optimum height of the telescope mirror M1 above the ground level for better seeing, relative contribution of the atmospheric seeing was measured using micro-thermal sensor pairs made from nickel wire of 25 μm diameter separated by a distance of 1 m and mounted at 6, 12 and 18 m above the ground on a 18 m-high mast on the hilltop. Accuracy of these measurements was 0.01°C. More technical details of the instrument and analyses of the data are provided elsewhere[19,20]. Based on 11 nights (over 8 h each) of simultaneous measurements of both micro-thermal fluctuations and DIMM seeing during March–April 1999, it was found that the major contribution to seeing degradation of 0″.86 comes from the 6–12 m slab of the atmosphere since it is located close to the ground and thermal gradients affect this layer most compared to the higher layers[20]. Analysis of data indicated that if M1 is located at a height of ~13 m above the ground level, one can achieve sub-arcsecond seeing for a significant fraction of observing time. Long-term atmospheric extinction behaviour of the Devasthal site has also been studied[21]. Table 2 lists the key characteristic parameters of the Devasthal site measured at optical wavelengths. The site offers over

**Table 2.** Characteristic parameters of the Devasthal site measured during 1996–2001

| Parameters | Value |
|---|---|
| Location | Altitude: 2424 ± 4 m; long: 79°41′04″E; lat: 29°21′40″N |
| Seeing (ground level) | 1″.1 (median); 0″.75 (10 percentile value) |
| Estimated seeing | 0″.86 for 6–12 m and 0″.22 for 12–18 m slabs above ground level |
| Wind | <3 m/s for 75% of time |
| Air temperature | 21.5°C to −4.5°C (variation during year) ≤ 2°C (variation during night) |
| Rain | 2 m (average over year, 80% during June–September) |
| Snowfall | 60 cm (average; during January and February only) |
| Clear nights | Out of 208 spectroscopic nights, 175 are photometric |
| Sky extinction (mag/air mass) | Average: $k_U = 0.49 \pm 0.09$; $k_B = 0.32 \pm 0.06$<br>$k_V = 0.21 \pm 0.05$; $k_R = 0.13 \pm 0.04$; $k_I = 0.08 \pm 0.04$<br>Best : $k_U = 0.40 \pm 0.01$; $k_B = 0.22 \pm 0.01$<br>$k_V = 0.12 \pm 0.01$; $k_R = 0.06 \pm 0.01$ |
| Relative humidity | ≤60% during spectroscopic nights; much higher during July–September |

200 spectroscopic nights in a year, out of which ~80% are of photometric quality. The Devasthal site became an Observatory in 2010, when the 1.3 m DFOT was successfully installed and commissioned[16,22]. It is worth mentioning here that till recently, there was no knowledge about near-infrared (NIR) properties of the site. Consequently, technical parameters specified for the telescope as well as for telescope-building were mainly based on the parameters estimated at optical wavelengths.

## Activities of the 3.6 m DOT facility

After financial approval in 2006, various activities of the 3.6 m DOT project gained momentum. Concrete planning was done and appropriate committees/teams were formed for timely execution of the project. The ensuing subsections describe all these developments in brief.

### *Specifications and design of the telescope system*

In June 2006, a meeting of leading Indian optical astronomers and experts in astronomical instrumentation was held at ARIES, Nainital, to discuss the key scientific programmes that can be carried out with a 4 m-class optical telescope, and accordingly optimize the telescope specifications. A detailed discussion on the availability of 4 m size mirror blanks with global potential supplier, led to freezing the size of M1 as 3.6 m. Based on scientific needs, two-mirror Ritchey–Chretein (RC) configuration with a field of view (FOV) of 30′ and effective focal ratio of 9 was decided for the 3.6 m DOT. The 2 m Himalayan Chandra Telescope at Hanle[14] has $f/9$ RC configuration, while the 2 m Inter-University Centre for Astronomy and Astrophysics (IUCAA) telescope at Girawali[23] has $f/10$ RC configuration. The choice of $f/9$ RC beam for the 3.6 m DOT can, therefore, provide easy interchangeability and testing of back-end instruments amongst the existing Indian large optical telescopes. The telescope has a Cassegrain focus where light beam can be directed towards either one axial port or one of the two side ports. The M1 is a Zerodur glass meniscus having 3.6 m optical diameter, *f*-ratio of 2 and 165 mm thickness, while M2 is a Astro-Sitall glass having 952 mm optical diameter, *f*-ratio of 2.5 and 120 mm thickness. Back focal distance of the telescope is 2.5 m.

The image profile of a celestial source obtained from an optical telescope is a convolution of optical quality of the telescope, seeing of the site, and structure of the imaged source. The telescope optical specification was therefore set to match the 10 percentile seeing of 0″.7 at Devasthal site and accordingly, the 80% encircled energy (E80) of a point source was set to <0″.45 for the entire FOV of 30′ at optical wavelengths. M1 was decided to be located at ~15 m above the ground, where contribution from ground level seeing is expected to be <0″.4. Furthermore, considering the science objectives related to high resolution optical spectroscopic study as well as for seeing limited imaging at the optical and NIR wavebands, it was decided to make use of modern, proven active optics technology, first time for an Indian optical telescope. Opto-mechanical design of the telescope as well as performance at AMOS are given elsewhere[10–12,24]. Figure 3 displays optical ray tracing diagram of both side ports and the main axial port. FOV of the 3.6 m DOT is 30′ diameter at the main axial port, whereas it is only 10′ diameter at both side ports. The specified pointing accuracy of the alt-azimuth mount telescope is <2″ RMS (root mean squared), the tracking accuracy of <0″.1 for both 1 min in open loop and 1 h in close loop. As the specified optical image quality (E80 < 0″.45 for $\lambda \leq 1.5$ μm) of the telescope is very sharp in comparison to seeing at the site, the wavefront error (WFE) of the system was measured using wavefront sensor (WFS) by averaging out seeing. Hence, a relationship between measured WFE and the E80 requirement was established by performing a Monte-Carlo analysis considering a large panel of WFE sources[25]. It was found that the specified optical image

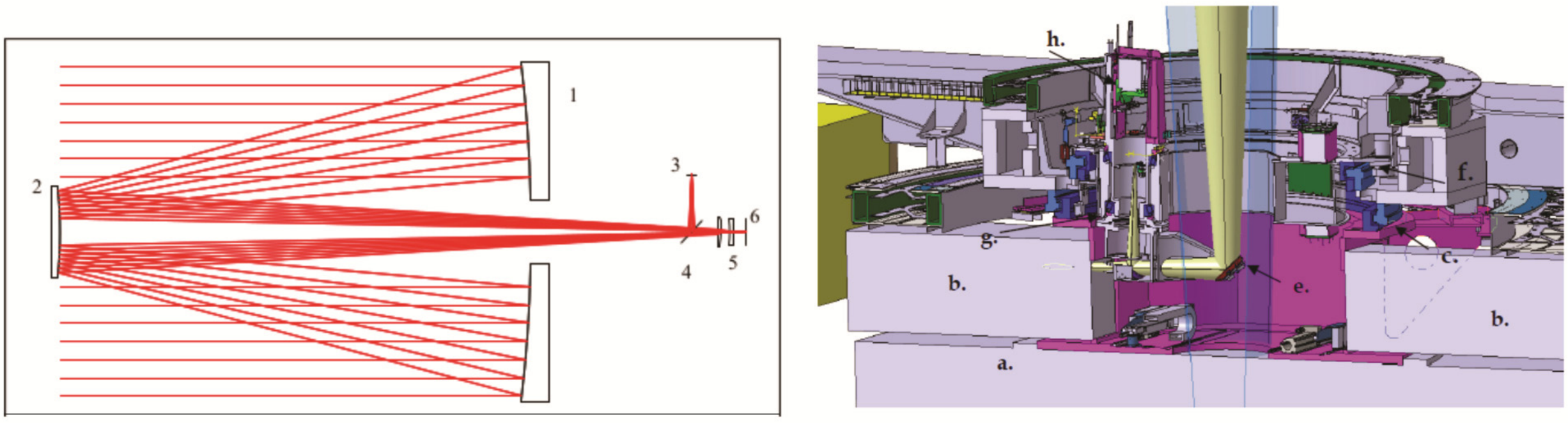

**Figure 3.** (Left) Ray tracing of the optical system of the telescope, with (1) the primary mirror, (2) secondary mirror, (3) side port focal plane, (4) side port folding mirror, (5) field corrector and (6) axial focal plane. (Right) Adapter Rotator Instrument Support Structure (ARISS): (a) Main instrument envelope. (b) side-port instrument envelope, (c) rotator bearing, (e) pick-off mirror, (f) adaptor bearing, (g) turn table and (h) optical bench with guider camera and wave front sensor.

quality of E80 < 0″.45 for $\lambda \leq 1.5$ μm could be reached for WFE < 0.21 μm.

Sophisticated techniques are employed for achieving and maintaining the image quality while the telescope is tracking objects in the sky. The 3.6 m DOT is equipped with an Active Optics System (AOS) that detects and corrects deformations, aberrations or any other phenomenon that degrade the image quality of the telescope[26,27]. The AOS compensates for distorting forces that change relatively slowly, roughly on timescales of seconds. It consists of a WFS, M1 mirror support system consisting of 69 actuators generating forces on M1 and three axial definers with load cells on M1, M2 mirror hexapod that supports M2 and the Telescope Control System (TCS) which acts as the interface between each element of the telescope. The sources of image quality degradation are due to inaccuracies in mirror manufacturing, mirror cell integration inaccuracies, gravity load, thermal effects and wind effects. A WFS is used to measure and analyse the wavefront coming from the telescope system. Focus, coma and tilt can be corrected with the M2 hexapod, whereas astigmatism, three-fold and spherical aberration are corrected with the actuators of M1 support. Load cells measure the residual forces on the three axial definers, and the actuators are used to keep these forces zero. The repeatable corrections on M1 and M2 mirrors are applied in open-loop mode (look-up table), whereas the close-loop mode applies both repeatable and non-repeatable corrections. The latter arises from thermal deformations and wind effects.

The Acquisition and Guiding Unit (AGU) is designed to measure both WFE and tracking errors during the operation of the telescope. A pick-off mirror can be aligned on a guide star located at the edge of the telescope's main axial port FOV in an annular region of 30′–35′. The guide star light beam is then directed towards an optical bench equipped with a WFS and a guide camera (Figure 3). The centroid of star on the guide camera system is used to correct the tracking errors, while the WFS camera system measures WFE.

TCS forms the interface between the hardware of telescope and the user. It consists of a weather station, mount control system and programmable logic controller. It provides access to both the operational and engineering control of the telescope hardware and also interfaces with the AOS, guiding unit system and the imaging instrument, if any. TCS is installed on a dedicated industrial PC with in-built GPS card and communication ports to communicate with the observatory control system, mount control system and the AOS. TCS runs under the LINUX platform within the Lab VIEW environment. It accepts coordinates of both the target and the guide star.

## *Telescope enclosure and dome control system*

The telescope enclosure building is designed to house a pier, the 3.6 m DOT and a mirror coating unit. Considering the limited space available at the site, a compact building has been designed. The specification and design of the telescope enclosure are given elsewhere[28]. The design and consultancy for the enclosure were done by the Precision Precast System (PPS), Pune, and construction of the enclosure was done by the Pedvak, Hyderabad. It is a building of great complexity involving structural (concrete and steel), mechanical and electrical engineering. It has three parts, viz. the rotating dome, the stationary dome-support structure and an auxiliary building. The 3.6 m DOT, designed to have natural frequency of 7.4 Hz, is mounted on a M25-grade concrete hollow cylindrical pier of 5 m inner diameter and 2 m wall thickness. The top of the pier is 8.26 m above the ground and its foundation is completely isolated from the telescope building. The natural frequency of the pier is designed to be 25.44 Hz, a value which is significantly different from the natural frequency of the telescope (7.4 Hz). Horizontal natural frequency of the self-loaded built pier was measured using 3C geophones and piezoelectric sensors. It has a mean value of 22 ± 2 Hz (ref. 29) and is consistent with the as-designed value. The rotating dome is a cylindrical, insulated structure with pitched roof having a diameter of 16.5 m and height of 13 m. It has 4.2 m wide opening slits and a wind screen. The telescope floor is 11 m above the ground level. In such a structure, heat trapped in the enclosures degrades the overall performance of the telescope. In order to minimize these thermal effects, the 3.6 m DOT enclosure is equipped with 12 ventilation fans which takes in air from the dome shutter and flushes it out through the dome vents. Three additional fans are provided at the enclosure basement in the prevailing wind direction for flushing the heat trapped in the telescope pier and in the technical room on the ground floor through underground ducts. These fans developed by NADI Airtechnics Pvt Ltd rotate at 1250 rpm, providing a flow rate of 7900 $m^3$/h. Before making observations, the dome shutter is opened and the ventilation fans are operated for a few hours and then stopped during the observations. In order to understand the effectiveness of dome ventilation system and to efficiently utilize it to reduce the effect of dome seeing, it is necessary to monitor the temperature gradient in the telescope and its enclosure. For this, a network of wireless sensors is being developed for deploying them near the heat load and in the wind path at different locations, so that the temperature variation of telescope structure, mirror, dome, pier, technical room, underground ducts, basement area and outside environment can be monitored[30]. The dome building is also equipped with a four-passenger lift which is used to transport instrument components from the ground to the telescope floor.

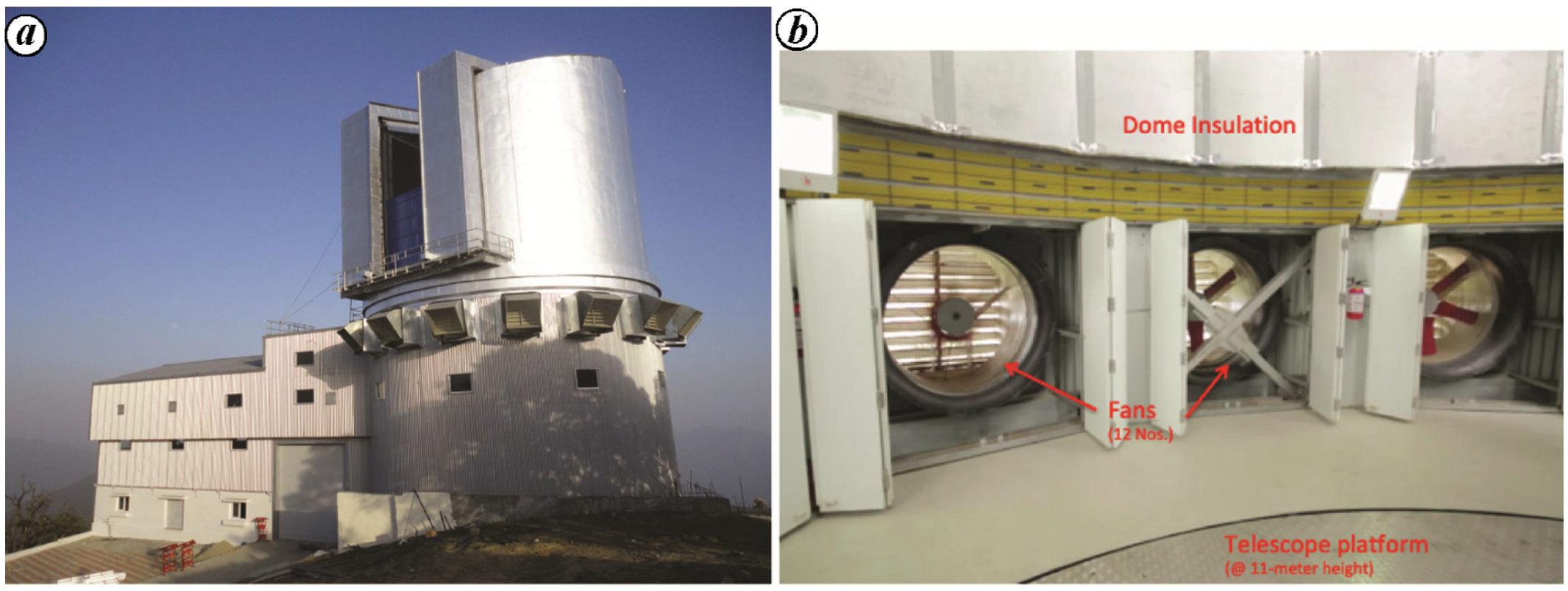

**Figure 4.** ***a***, Photograph of the 3.6 m telescope enclosure at Devasthal. The wind screen is seen in blue colour. The big square-size shaped louvres are hoods of 12 ventilation fans. ***b***, A small portion of dome insulation, a few ventilation fans in the dome and a tiny part of the telescope platform. The fans suck outside air through the dome slit and other openings in the dome building, and exhaust to the outside.

The telescope building was also designed to house overhead cranes so that the integration of the entire telescope could be made from within the building by lifting components through its hatch. For this, two single-girder cranes in the extension building and two under-slung cranes in the dome of 10 MT capacity each were specifically designed and developed indigenously and used successfully during installation of the telescope. Further technical details of the cranes can be found elsewhere[31].

A cable anti-twister has been designed and installed in the pier to facilitate delicate instrument cables through the hole provided in the telescope pier. The critical cables such as helium supply, cryoline and optical fibre used in back-end instruments will be routed through the anti-twister. Figure 4 shows photographs of the as-built building along with details of dome insulation and fans.

The centre of the 3.6 m DOT is 1.85 m away from the dome centre at an angle of 255° with respect to the north direction. The dome slit motion is synchronized with that of the telescope using an algorithm developed in-house in Python language and implemented using the dome control system (DCS). An absolute multi-turn encoder is mounted with one of the dome wheels to fetch dome azimuth position. A photoelectric sensor is also mounted to get the homing position and error correction for dome location. A micro-controller-based interface card is used to interface hardware and DCS using ethernet converter. DCS acquires the position of azimuth each second from TCS. The dome can be rotated in both fast (495 sec/rotation) and slow (770 sec/rotation) modes. Dome motion tracks telescope motion with an accuracy of ~40″. The margins of error allowed by the dome slit width range between ~1° and 1.7° depending upon the location of M1 during operation of the telescope. The complete description of the DCS is given elsewhere[32]. Figure 5 shows a snapshot of the DCS graphics user interface (GUI) window and fully aligned dome and telescope.

## *Aluminium coating unit*

As the reflectivity of mirrors used in the telescope decreases due to weathering, their realuminization is necessary at regular intervals. In order to optimize cost of the 3.6 m DOT project, its mirror M1 was transported to Devasthal without aluminium coating. An aluminizing coating unit was therefore designed and developed indigenously and installed in the extension building of the telescope by Hind High Vacuum, Bengaluru. Both mirrors M1 and M2 of the telescope can be unmounted and brought down to the coating facility using overhead cranes installed in the building. The magnetron sputtering technique is used for the coating[33]. The facility consists of a vacuum chamber made of stainless steel. A rotary pump, roots pump and two cryo pumps are connected to the chamber through valves to evacuate the chamber up to the level of one millionth of a millibar. The chamber takes about 6 h to achieve the required vacuum level. A mirror washing unit is also installed.

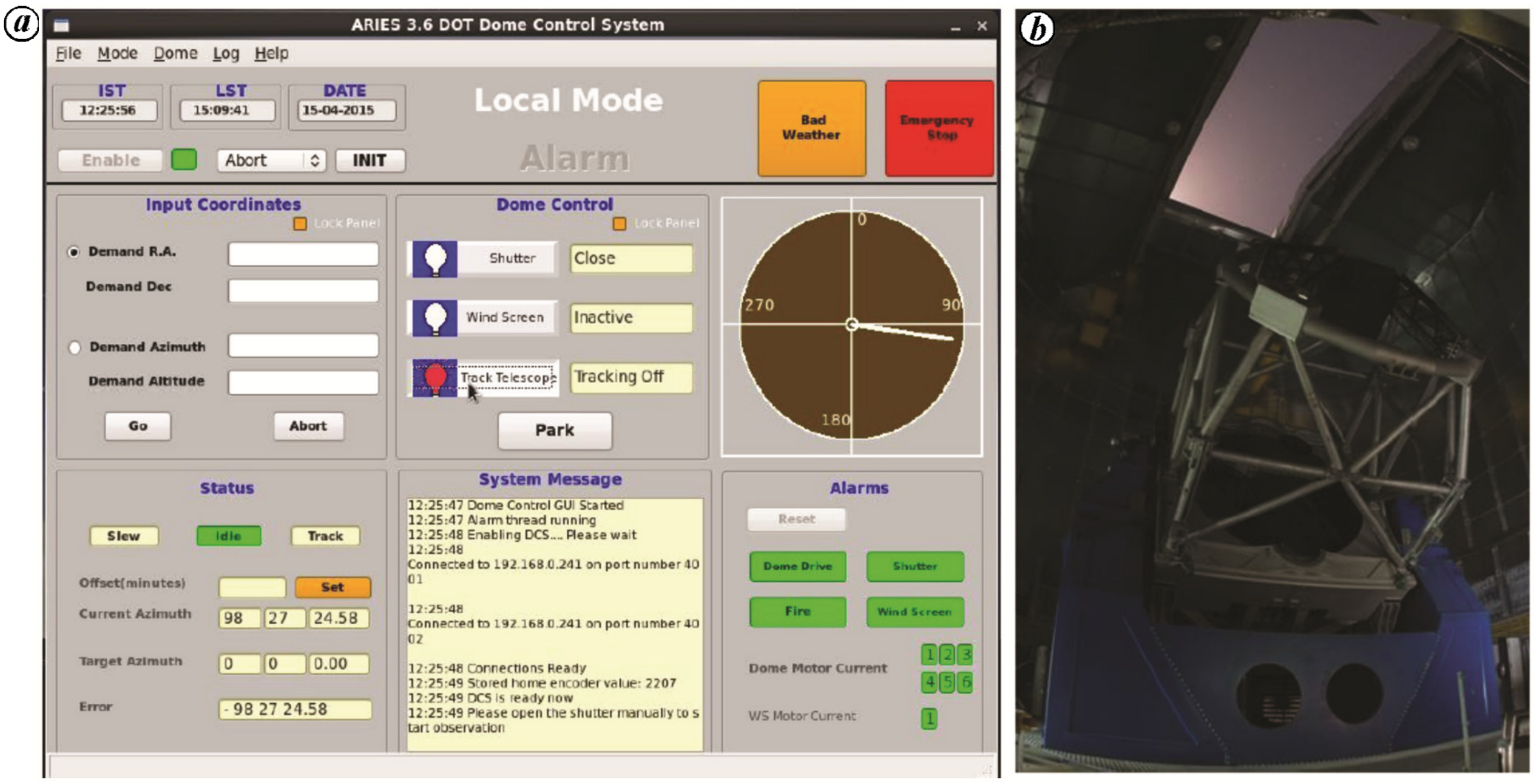

**Figure 5.** The software used for synchronization of telescope motion with that of the off-centred dome, designed and developed by ARIES. (***a***) DCS graphics user interface (GUI) and (***b***) telescope in synchronized position with dome to the right.

A detailed description of the coating plant installation, cleaning and coating procedures for M1 and testing results of the samples is provided elsewhere[34]. M1 was first cleaned and then coated with reflective aluminium in February 2015, using the above described coating unit at Devasthal (Figure 6). A mean reflectivity of ~86% was achieved in the $\lambda$ range 0.4–1 μm. M2 was coated with aluminium along with a protective layer deposition of $SiO_2$ in September 2010 at LZOS, and found to have a mean reflectivity of 87% in the $\lambda$ range 0.4–1 μm. Figure 6 shows the reflectively of freshly coated mirrors M1 and M2. So far, this facility has been used thrice for aluminium coating of M1 including the last one carried out during September 2018.

## *Installation and sky performance of the 3.6 m DOT*

The mechanical integration of the telescope was achieved using overhead cranes installed in the telescope building. These cranes can lift the telescope parts with hoist speed as slow as 3 mm/sec, and they also have provision to move in tandem. Parts of the 3.6 m DOT were transported in a number of boxes/containers and the maximum weight of a single part was 14 MT. The whole telescope of 150 MT was installed successfully on top of the telescope pier using these overhead cranes. The freshly coated mirror M1 was integrated with the telescope in January 2015. The 3.6 m DOT was fully assembled in March 2015 and its engineering verification tests were performed till May 2015, while rigorous on-sky tests, namely pointing accuracy, tracking accuracy, optical quality and AGU guider sensitivity were performed during October 2015 to February 2016. For this, four instruments, namely test-camera, test-WFS, AGU-camera and AGU–WFS were used. An air-cooled Micro-line ML 402ME CCD having a pixel size of 9 μm and chip size of $768 \times 512$ pixels was used for both test-camera and AGU-camera having a plate scale of 0″.06/pixel and 0″.166/pixel respectively. The Micro-line ML4710-1-MB with $1024 \times 1024$ size of 13 μm pixels was used for both test-WFS and AGU-WFS having lens-let array of $33 \times 33$ and $11 \times 11$ respectively. The test-WFS at the Cassegrain axial port is used to measure image quality of the actively corrected telescope. A comprehensive set of data were collected, analysed and the test procedures and results were published as a professional technical report[25]. Table 3 lists as-built specifications. A total of six measurements of optical quality of the telescope resulted in E50 of 0″.15 ± 0″.03, E80 of 0″.26 ± 0″.04 and E90 of 0″.37 ± 0″.07 (ref. 25). They are better than the corresponding tendered specifications of 0″.3, 0″.45 and 0″.6 respectively.

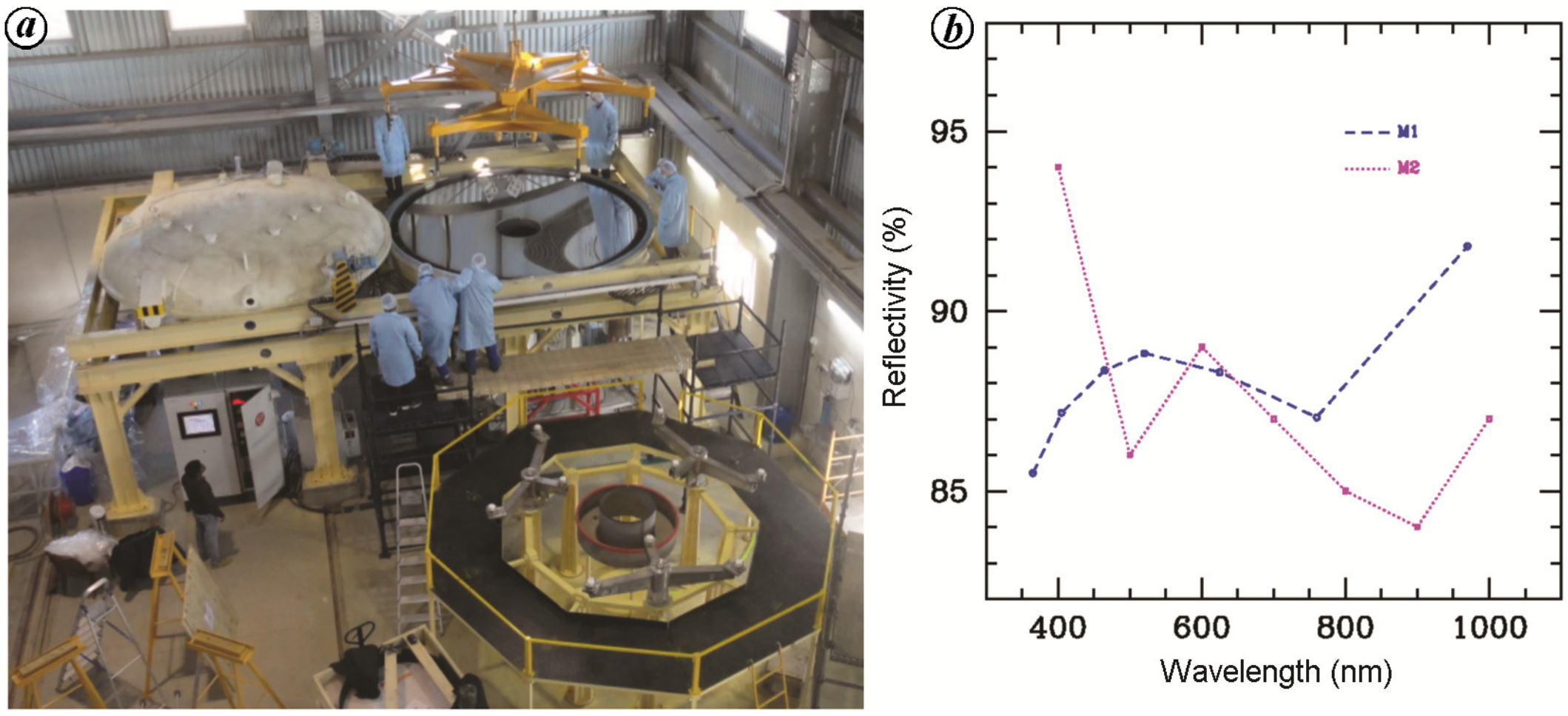

**Figure 6.** ***a***, Aluminium coating plant at Devasthal. The washing unit with nine whiffle tree support points can be seen in the foreground. The mirror M1 resides in the coating chamber. The mirror handling tool is seen just above M1 in yellow colour. ***b***, Reflectivity curves of both mirrors M1 and M2.

**Table 3.** As-built key characteristics of the 3.6 m DOT

| Parameters | Value |
|---|---|
| Pointing accuracy | Specified RMS value: <2″ and measured RMS values: 1″.1, 1″.3 and 1″.2 for side port-1, side port-2 and main port respectively. |
| Tracking accuracy open loop (without guider) | Specified RMS for 1 min: <0″.1; Measured (″): 0.06, 0.08, 0.08, 0.08, 0.09, 0.09, 0.10, 0.12, 0.12, 0.14; mean : 0″.08 ± 0″.03; specified peak RMS for 15 min: <0″.5; measured (″): 0.16, 0.22, 0.24, 0.24, 0.25, 0.32, 0.34, 0.35, 0.36, 0.42. |
| Tracking accuracy close loop (with guider): | Specified RMS for 1 h: < 0″.11; measured (″): 0.07, 0.07, 0.08, 0.09, 0.09, 0.09, 0.10, 0.10, 0.10, 0.11, 0.13, 0.13; mean : 0″.09 ± 0″.02. |
| Optical image quality | Specified (diameter): E50 < 0″.3; E80 < 0″.45; E90 < 0″.6; measured (″): E50 values are 0.19, 0.13, 0.13, 0.12, 0.18, 0.14; E80 values are 0.37, 0.27, 0.23, 0.25, 0.33, 0.23; E90 values are 0.49, 0.36, 0.31, 0.32, 0.42, 0.30. Note: E50, E80 and E90 values are inferred from analysis of aberrations of optics with a mean of 0″.15, 0″.20 and 0″.37 respectively. |
| AGU guider sensitivity | Specified: *V* mag of ~13 mag; measured: 12.85 mag star tested successfully. |

The point spread function (PSF) of a star has three main contributions, viz. seeing, tracking and optical quality. It has been observed that the optical quality of the telescope can be maintained at the level E80 of 0″.26 ± 0″.04 and this corresponds to a FWHM value of 0″.17 ± 0″.03. As M1 of the telescope is ~15 m above the ground level, ~0″.23 FWHM can be expected under best seeing conditions. Hence, considering the tracking accuracy of 0″.1 RMS, it is expected that the telescope can deliver FWHM images of stellar sources to ~0″.3 in best seeing conditions. The effect of local seeing on the measurement of WFE was observed on 3 November 2015. The ventilation fans were switched off at the start of the night and it was observed that the WFE recorded using both test-WFS and AGU-WFS did not converge, and its value could fluctuate around 350 nm. However, the WFE converged to values <200 nm within 1 h of starting the dome ventilation fans. A gain of ~100 nm was thus measured in test-WFS, which can be attributed to the effects of local temperature gradients. The sensitivity of AGU-WFS has been found to be satisfactory for a star of $V = 12.85$ mag and $(B – V) = 0.51$ mag with an integration time of 300 sec.

We observed double stars with known angular separation during November–December 2015 using the test-camera. Binary stars having separation of sub-arc-second were clearly resolved (Figure 7). In one of the best observations, a binary star with known angular separation of ~0″.4 was well resolved on the night of 30 November 2015.

## Cassegrain port instruments

The telescope provides three Cassegrain ports for mounting instruments (Figure 8). The main axial port is designed for mounting instruments weighing 2000 kg. Any axial port instrument can be mounted with the telescope interface plate (TIP) having a diameter of 1.5 m and thickness of 3 cm, and it can occupy a space of 1.8 m length below TIP in the shape of a cylindrical (0.8 m length)-cum-conical (1 m length) shape. From TIP to 0.8 m length, it has a shape of 3 m diameter cylinder, while from 0.8 to 1.8 m below TIP, it has the shape of a cone with diameter decreasing from 3 to 1 m. The centre of gravity (CG) of the axial port instrument is 0.8 m below TIP. The positioning of CG for any instrument is critical, since a shift of 1 cm of CG position in the vertical axis will create a torque of 200 Nm, which is close to the altitude-bearing friction. The telescope has the capacity to bear imbalance of 2000 Nm on altitude axes and 400 Nm on azimuth axes. Imbalances are adjustable using motorized weights on altitude axes and fixed weights on azimuth axes. The side port instruments can have a weight of 250 kg each with CG of 0.62 m away from the interface plate. Further technical details are given elsewhere[3]. The main instrument and the two side ports are fixed to a structure, called Adapter Rotator Instrument Support Structure (ARISS), including a device which rotates the image of the sky. The Side-Port Fold Mirror is also located in this structure. All these units are nested at the rear of M1, just above the instrument.

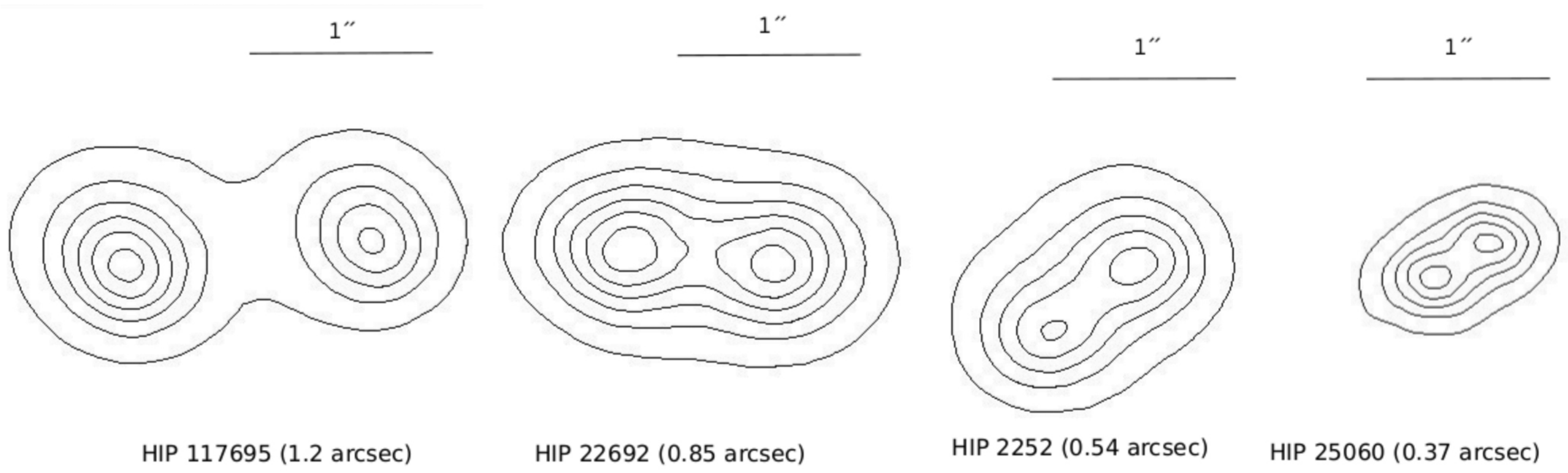

**Figure 7.** Iso-intensity contour images of four binary stars. These images were taken with the test camera mounted at the axial Cassegrain port of the 3.6 m DOT. Angular separation between binary stars ranges from 0″.37 to 1″.2.

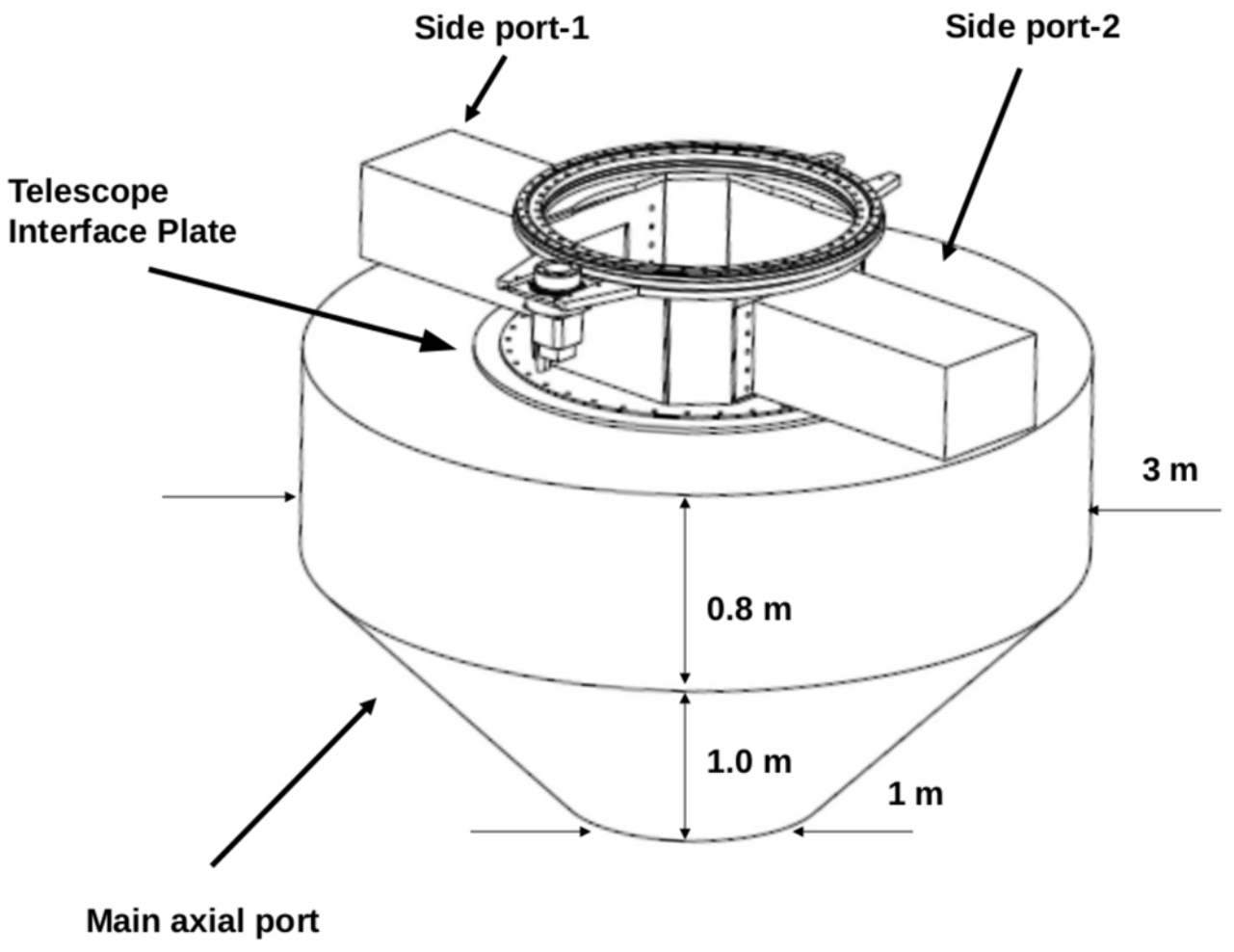

**Figure 8.** A sketch map of the focal planes showing the main axial port and both side ports.

Figure 9 shows photographs of the as-built telescope along with back-end instruments being used for regular observations. The 4K × 4K CCD IMAGER and ARIES-Devasthal Faint Object Spectrograph and Camera (ADFOSC) are used at optical wavelengths, while TIRCAM2 (acronym for TIFR NIR Imaging Camera-II), is used in the NIR region. Technical information on these back-end instruments is given in the following subsections.

## *IMAGER: 4K × 4K CCD camera*

The IMAGER is indigenously designed and assembled at ARIES. It has wavelength sensitivity between 0.35 and 0.9 μm. Technical details of optical and mechanical design, motorized filter wheels and data acquisition system of the instrument along with its expected performance on the 3.6 m DOT under different seeing conditions are given elsewhere[35]. The pixel size of a blue-enhanced, liquid nitrogen-cooled (about – 120°C) STA4150 4K × 4K CCD sensor is 15 μm square and presently, it is equipped with standard broadband Bessel *U*, *B*, *V*, *R*, and *I* and SDSS *u*, *g*, *r*, *i*, and *z* filters and covers FOV of 6′.5 × 6′.5 when mounted on the telescope. The IMAGER has been mounted a number of times on the axial port of the telescope since the year 2015 for testing, calibration and science observations. The images of extended sources, viz. galaxies and star clusters were recorded[35]. A few published results based on observations of this instrument are given in the next section (Figure 10). In future, it may be permanently mounted at one of the side ports of the telescope.

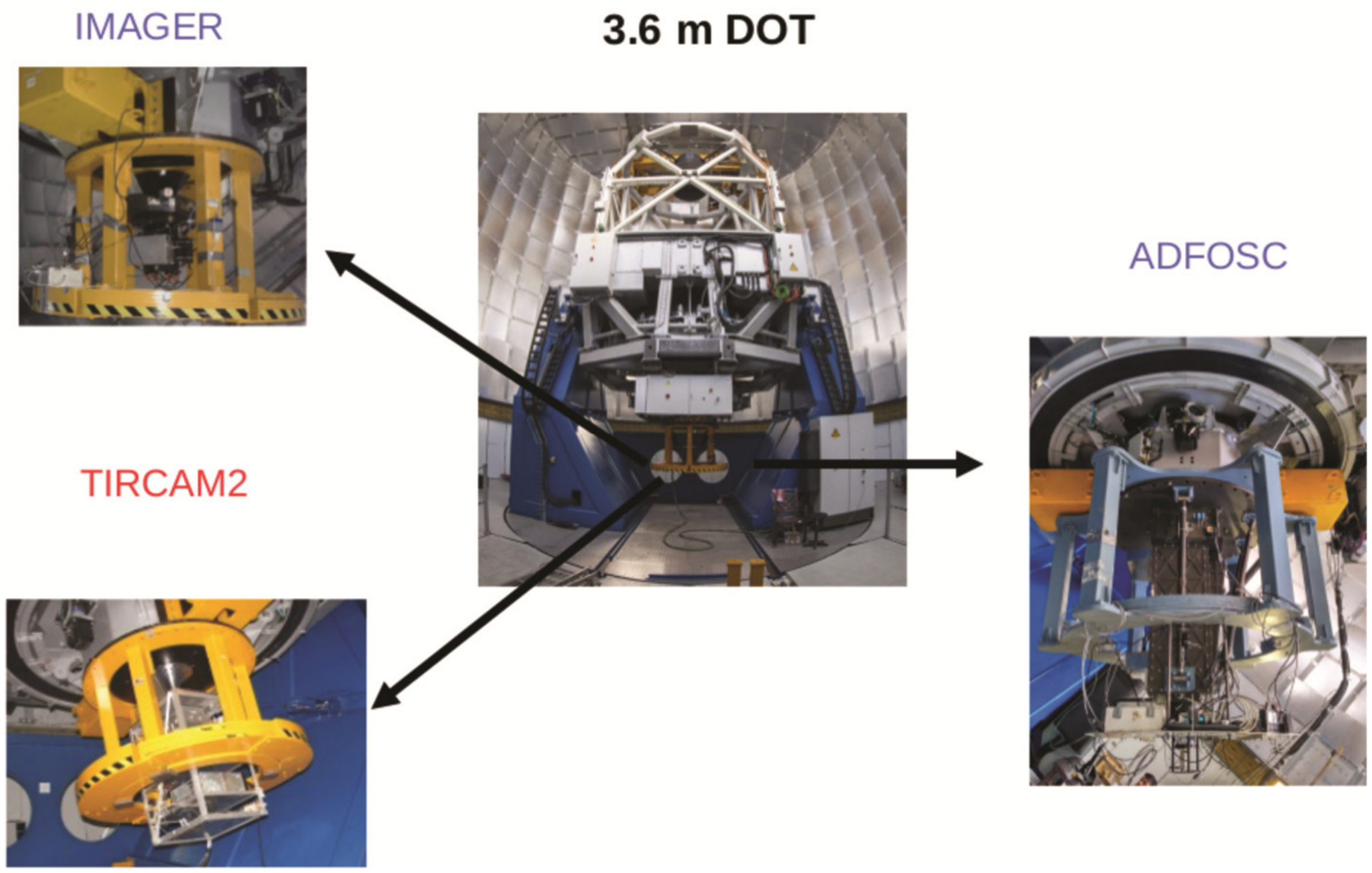

**Figure 9.** The as-built 3.6 m DOT (centre). Photographs of back-end instruments 4K × 4K CCD IMAGER, ADFOSC and TIRCAM2 used for regular observations are also shown.

## *ADFOSC*

The ADFOSC is a complex instrument with a low-resolution slit spectrograph-cum-imager having wavelength sensitivity between 0.35 and 0.9 μm. It has been designed, developed and assembled in-house at ARIES with national and international collaboration. After mounting on the main axial port of the telescope, it has been tested and also used for science observations since 2016. Its optical design consists of a collimator and a focal reducer converting the telescope $f/9$ beam to $f/4.3$ beam. The $\lambda$ dependence of the ADFOSC transmission optics is estimated to be 84%, 84%, 86%, 87%, 80% and 68% at $\lambda = 0.36$, 0.464, 0.612, 0.744, 0.89 and 1 μm respectively[36]. To record images from the instrument, a closed-cycle cryogenically cooled, grade-0, back-illuminated E2V 231-84 chip of 4096 × 4096 square pixel CCD camera having an image area of 61.4 × 61.4 sq. mm was also designed, developed and assembled in-house. The grisms having transmission gratings with 300–600 lines/mm are mounted at the pupil plane. The spectral resolution with 1 arc sec of atmospheric seeing is in the range 500–1000. An 11° wedge prism is also available to obtain very low dispersion spectrum suitable for slit-less spectroscopy. Further technical details and optical design of the instrument are given elsewhere[37,38]. ADFOSC is a versatile instrument since it can be used in a number of modes, namely (a) broad and narrow band photometric imaging, (b) long-slit low-resolution ($\lambda/\Delta\lambda < 2000$) and slit-less spectroscopy, and (c) fast imaging (up to millisecond cadence). Presently, it is equipped with SDSS *u*, *g*, *r*, *i*, and *z* filters, 8′ long slits, grisms and narrow-band filters, and can image a FOV of 13′.6 × 13′.6 when mounted on the telescope. ADFOSC has been used in both imaging and spectroscopy modes for the observations of stars, ionized star-forming regions and galaxies[38]. The expected efficiencies of ADFOSC in the imaging mode are estimated to be 9.6%, 33.7%, 42.1%, 45.3% and 27% at $\lambda = 360$, 464, 612, 744 and 890 nm respectively[36]. The corresponding values of spectral resolution ($\Delta\lambda$) are 28, 64, 58, 61 and 61 nm respectively. Its efficiency at the CCD response cut-off at $\lambda = 1000$ nm is estimated to be only 4%. In spectroscopic mode, the sensitivity of ADFOSC can be judged from the spectrum of a $V \sim 15$ mag bright SN 2017 gmr taken on 17 November 2017 in a single exposure of 5 min (Figure 11). The average value of spectral resolution ($\Delta\lambda$) is ~1 nm. The values of RMS noise ($10^{-15}$ erg/cm$^2$/s/nm) are 4, 2, 0.8 and 1 at $\lambda = 450$, 600, 750 and 900 nm respectively. In Figure 11, the signal-to-noise (S/N) ratio of the $H_\alpha$ and $H_\beta$ lines is 70 and 20 respectively.

## *TIRCAM2*

TIRCAM2 is a closed-cycle helium cryo-cooled 512 × 512 pixel imaging camera with wavelength sensitivity from 1 to 5 μm. Pixel scale of the camera on the telescope is 0″.17. TIRCAM2 has been mounted on the axial port of the telescope for tests, characterization and science observations since 2016. It is equipped with standard *J* (1.2 μm), *H*(1.65 μm) and *K*(2.19 μm) broad ($\Delta\lambda \sim 0.3$–0.4 μm) photometric bands, and narrow ($\Delta\lambda \sim 0.03$–0.07 μm) band $B_{r-\gamma}$ (2.16 μm); $K_{\text{cont}}$ (2.17 μm); polycyclic aromatic hydrocarbon (PAH; 3.29 μm) and narrow band L (nbL; 3.59 μm) filters. Observations taken with this instrument mounted on the telescope show that the camera is capable of observing sources up to 19.0, 18.8 and 18.0 mag with 10% photometric accuracy in *J*, *H* and *K* bands respectively, with corresponding effective exposure times of 550, 550 and 1000 sec respectively. It is also capable of detecting the nbL band sources brighter than ~9.2 mag and strong (≥0.4 Jy) PAH-emitting sources like Sh 2-61. TIRCAM2 on the 3.6 m DOT is well suited for the study of low and very low mass stellar populations (M-dwarfs, brown dwarfs), strong mass-losing stars on the asymptotic giant branch and young stellar objects still in their proto-stellar envelopes. The technical and performance details of this instrument are given elsewhere[39,40].

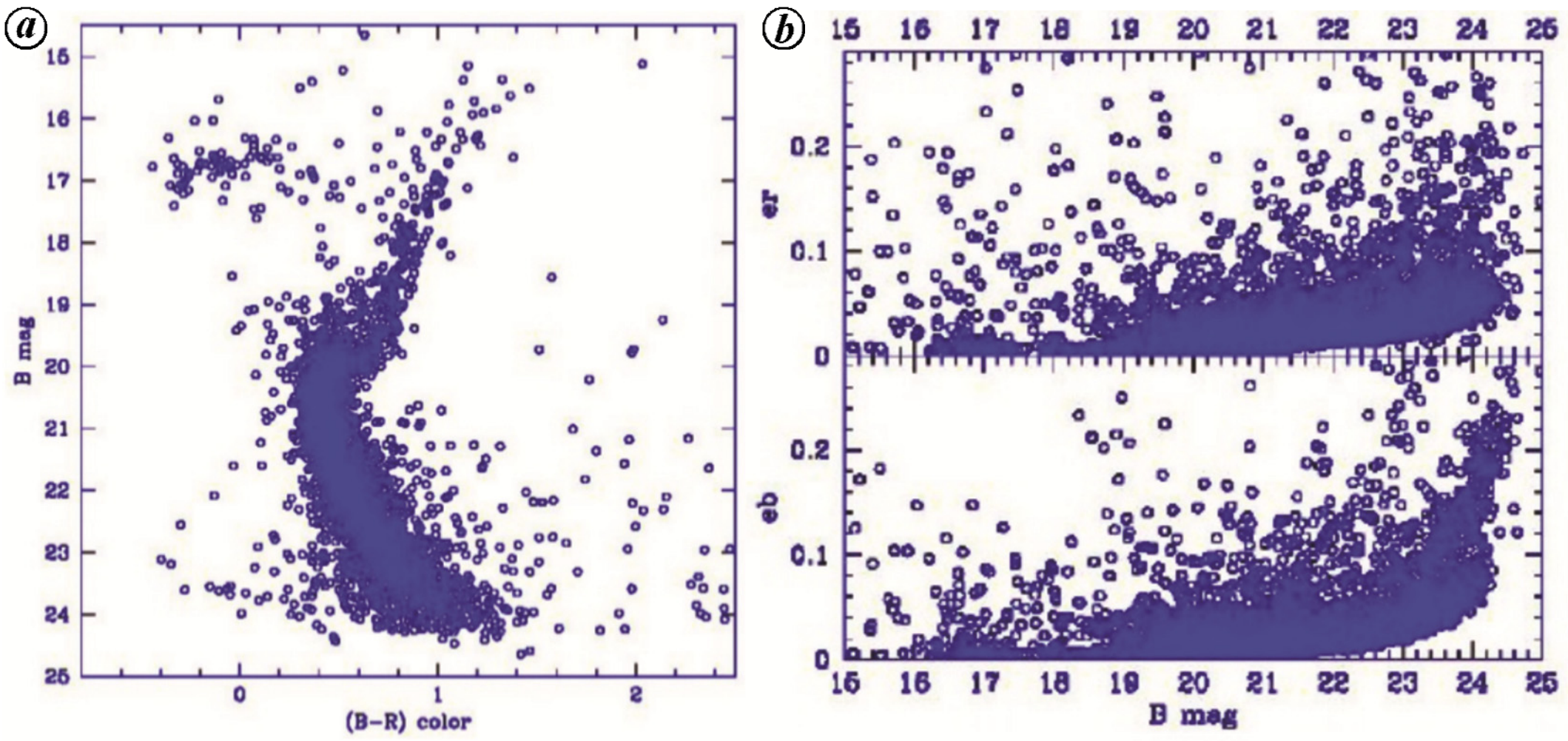

**Figure 10.** ***a***, *B* versus (*B–R*) CMD of the galactic globular cluster NGC 4147. ***b***, Photometric errors in magnitude as a function of *B* magnitude.

## Highlights of recent results from the 3.6 m DOT

After successful installation and performance verification, technical activation of the 3.6 m DOT was done jointly by both the Indian and Belgian Premiers on 30 March 2016 from Brussels, Belgium. Scientific potential and capabilities of the telescope are published elsewhere[1,41,42]. Observing proposals from the users are submitted online twice in a year for both observing cycles, namely cycle-A (February to May) and cycle-B (October to January). The website (http://www.aries.res.in/dot) provides relevant information regarding the policies and procedures followed in the allotment of observing time for the telescope. Based on scientific merit of the submitted proposals, the Belgian and Indian time-allocation committees allot observing time to the proposers of their countries. Presently, 33% and 7% observing time has been allotted to the proposers from ARIES and Belgium respectively, and the remaining 60% observing time to other proposers. During cycles 2017A, 2017B and 2018A, 120 research proposals were received with an over-subscription factor of ~3. Only 41 proposals were allotted telescope time.

After technical activation on 30 March 2016, the telescope is in regular use. There were 550 useful/spectroscopic nights (with a minimum period of 4 h of clear sky observations) during April 2016 to December 2018. About 80% of them were of photometric quality. The number of nights lost due to bad weather and monsoon (June to September) was 155, 151 and 159 in 2016, 2017 and 2018 respectively. These numbers are as expected based on the site survey results given in Table 2. About 40% of the useful nights (mostly during 2016 and 2017) were used for instrument testing, verification and good-quality science observations of 41 proposals approved by the time-allocation committees. Back-end instruments mounted on the telescope were the CCD IMAGER, ADFOSC and TIRCAM2 for 80, 75 and 65 useful nights respectively. Another 20% of useful nights were used for the training of scientists and engineers of ARIES, testing and routine maintenance of the telescope, including major refurbishment activities such as replacement of guider-arm ball screw, the ARISS umac and azimuth drive, etc. Remaining 40% of useful nights (mostly in 2018) could not be utilized due to technical faults in critical parts of the telescope, namely azimuth motor on 23 November 2017, ungluing of a few pads supporting mirrors M1 and M2 and failure of the WFS shutter; remedial actions were taken. Presently the telescope is functional with reduced azimuth slewing speed of

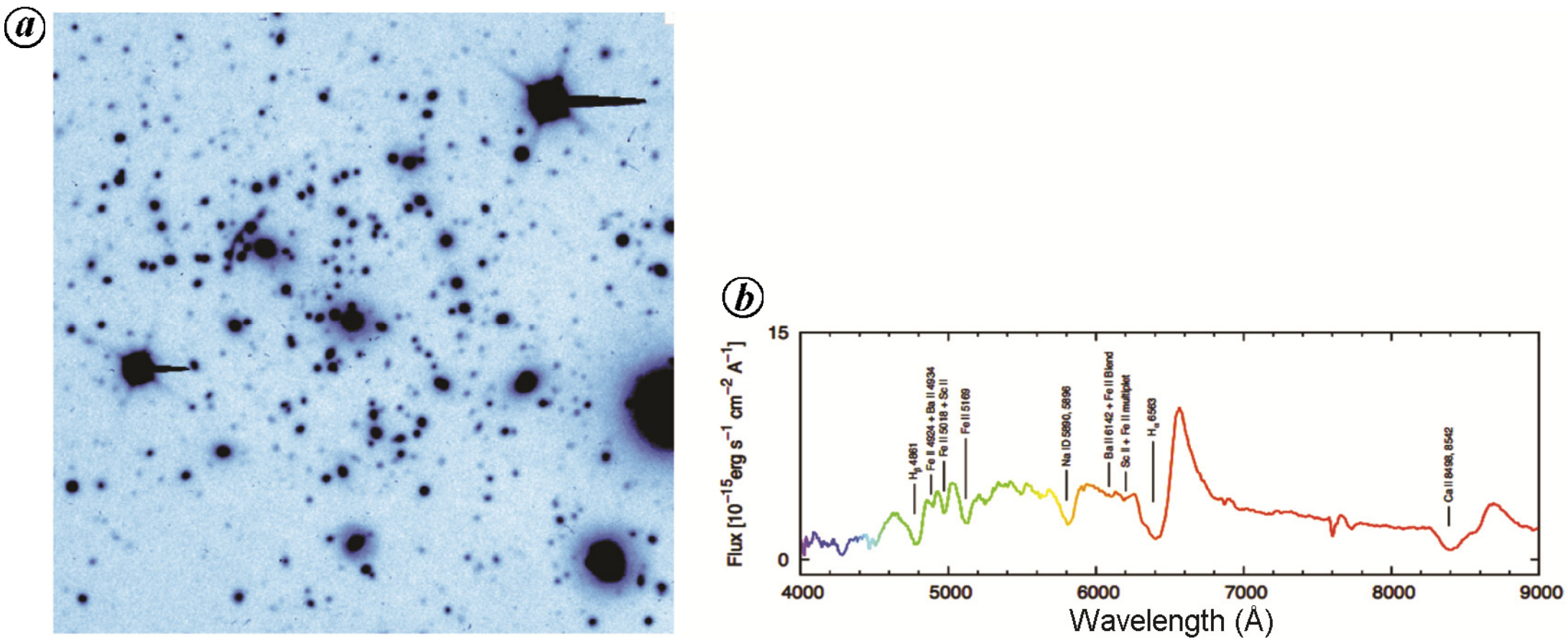

**Figure 11.** ***a***, Calibrated image of central region (~3′ × 3′) of the galaxy cluster Abell 370 taken in *i*-band with ADFOSC in the imaging mode mounted on the 3.6 m DOT. ***b***, Calibrated spectrum of SN 2017gmr extracted from a single exposure of 5 min, obtained in the spectrum mode of ADFOSC.

~1°/sec. It is expected that the azimuth motor will be replaced soon so that normal functioning of the telescope is restored.

The observed science proposals cover astrophysical topics related to studies of star formation, star clusters, stellar evolution, optical transient objects, variable stars, exo-planets, distant galaxies, AGN and quasars. In the following sub-sections, a few published science results are summarized along with performance of the telescope at optical and NIR wavelengths.

### *Colour–magnitude diagram and search of variables in the star cluster*

It is well known that the colour–magnitude diagram (CMD) of a star cluster is an extremely valuable tool for the study of stellar evolution. Figure 10 *a* shows the CMD of the galactic globular cluster NGC 4147 observed with the 3.6 m DOT on 23 March 2017 during a dark night. The $B$ and $R$ broadband images have exposure time of 20 min each, but have different average FWHM values of 1″.20 and 1″.12 respectively. The sky brightness in $B$ and $R$ bands was estimated to be $22.29 \pm 0.34$ and $19.36 \pm 0.21$ mag/arc sec$^2$ respectively. Under these sky conditions, it can be seen from the magnitude–error diagram (Figure 10 *b*) that stars of $B = 24.5$ and $R = 23.5$ mag have been detected with S/N of 5 at an exposure time of 20 min (ref. 35). A well-defined cluster main sequence up to ~4 mag below the turn-off point is clearly visible in the $B$ versus $(B - R)$ CMD along with well-developed red-giant and blue horizontal branches. Some blue straggler stars (BSS) are also located above the turn-off point.

A search for variable stars has been carried out based on 339 and 302 CCD images of NGC 4147 taken in $V$ and $R$ filters with the CCD IMAGER mounted on the 3.6 m DOT during March–April 2017 (ref. 43). The FWHM values of these images ranged from 0″.7 to 1″.0. These observations identified 42 periodic stellar variables. Among them, 28 are newly discovered. Their location in the $V$ vesus $(V - R)$ CMD and variability characteristics indicates that the number of new RRc, EA/E, EW and SX Phe type variables is 7, 8, 5 and 1 respectively. The distance derived using light curves of RRab stars is consistent with that obtained from the observed $V$ versus $(V - R)$ CMD.

### *Optical detection of a GMRT-detected radio galaxy*

A high redshift candidate ($z \sim 4.8 \pm 2$) radio galaxy source designated as TGSS J1054 + 5832 was identified at 150 MHz using the GMRT images. In 2018, optical observations of this source were obtained in SDSS $r$ and $i$ pass bands with the 3.6 m DOT using ADFOSC. No clear detection of the source is made in the SDSS $r$-band, but a statistically significant optical detection in SDSS $i$-band has been made for this steep spectrum candidate high redshift galaxy[36]. The $i$-band AB magnitude for the source is estimated to be $24.3 \pm 0.2$.

### *Images of the galaxy cluster Abell 370*

Figure 11 shows an $i$-band calibrated image of the central region (~3′ × 3′) of the galaxy cluster Abell 370 taken with the 3.6 m DOT in the imaging mode of ADFOSC. Most of the diffuse objects in the image are clustered galaxies. The famous prominent giant gravitational-lens arc is easily visible. The sources as faint as $i \sim 24$–25 mag are detected. Location of these sources matches well with those present in the deep image of the same region available from the Hubble Space Telescope Legacy Program[38]. It indicates that angular resolution of sub-arcsecond has been achieved in the image. A detection sensitivity of ~24.5 mag in 1 h of integration time and sky brightness of ~20 mag/arcsec$^2$ in the $i$-band have been obtained with a photometric precision of 0.2 mag (ref. 38).

### *Optical observations of GRB afterglows and supernovae*

Optical observations of GRB 130603B afterglow and its host galaxy have been obtained in the $B$ and $R_c$ broadband filter with the 4K × 4K CCD IMAGER mounted on the axial port of the 3.6 m DOT. Exposure time in each filter was 600 sec. The estimated $B$ and $R_c$ magnitudes of the host galaxy were 22.13 ± 0.05 and 20.72 ± 0.02 respectively. These data have been used to construct the multi-wavelength spectral energy distribution of the host galaxy[44]. It indicates that the host galaxy is young and blue with moderate values of star-formation activities. The observations used here were obtained in 2017 after 1387 days of the GRB event.

The 4K × 4K CCD camera mounted on the axial port of the 3.6 m DOT was used to obtain $R$-band images of the Type II supernova ASASSN-16ab/SN 2016B located in the galaxy PGC 037392 (ref. 45). The images were taken on 2 April 2017, 465 days after explosion of the SN 2016B, in 2 × 2 binning mode of the CCD camera. The estimated value of $R$ was 19.79 ± 0.05 mag.

The optical spectrum of ~73 day-old supernova (SN) 2017gmr at $V \sim 15$ mag was taken with ADFOSC on 17 November 2017 at an exposure time of 5 min (ref. 38). It covers the $\lambda$ range from 0.4 to 0.9 μm (Figure 11). The line strengths of the prominent emission lines of the elements H, C, N, O, Fe were detected. It is a hydrogen-rich (Type-II) supernova and most likely produced from the explosion of a runaway massive star. The line features and strengths of the SN 2017 gmr spectrum have excellent correspondence with those of archived spectra of similar types of supernovae. The supernovae at this stage evolve rapidly and the line features change every few days. The spectroscopic observations of supernovae over regular intervals provide crucial information on evolution and nucleo-synthesis of various elements in the Universe.

### *Fast imaging of the crab pulsar*

The Crab pulsar (PSR B0531+21), a young neutron star spinning rapidly with a rotation period of 33.7 ms, was observed with the 3.6 m DOT using fast-imaging mode of ADFOSC with a frame-transfer and electron-multiplying CCD camera[38]. An effective cadence of 3.37 ms was achieved during observations. Repeating optical pulses from the Crab pulsar were detected.

### *Sky performance at optical and NIR wavelengths*

The above-mentioned optical observations indicate that the night-sky brightness values at Devasthal are 22.29 ± 0.34 and 19.36 ± 0.21 mag/arcsec$^2$ in Bessel $B$ and $R$ bands respectively, while they are ~21 and ~20.4 mag/arcsec$^2$ in the SDSS $r$- and $i$-bands respectively. Stars of $B = 24.5$ and $R = 23.5$ mag have been detected with S/N of 5 at an exposure time of 20 min and a FWHM of ~1″.2. A distant galaxy of 24.3 ± 0.2 mag has been detected in the SDSS $i$-band in 1 h of exposure time and FWHM ~1″.5. These numbers indicate that the night sky at Devasthal is dark and it has not degraded since late 1990s when the site characterization was carried out. During good-sky conditions, Baug *et al.*[40] estimated 16.4, 14.0, 12.2 and 3.0 mag/arcsec$^2$ as sky brightness in $J$, $H$, $K$ and $nbL$ bands respectively, which are comparable with other observatories like Hanle[14], Calar Alto Observatory[46] and Las Campanas Observatory[47]. The Devasthal sky performance at NIR wavelengths has been characterized in this study. In the $K$-band, stellar images with FWHM of ~0″.6–0″.9 were observed routinely, while the best value of FWHM (~0″.45) was observed on 16 October 2017. The 3.6 m DOT is therefore adequate for deep optical and NIR observations that are comparable to other 4 m-class telescopes available worldwide.

## Summary and future outlook

The on-sky performance of the 3.6 m DOT reveals that the quality of its optics is excellent and capable of providing images of the celestial bodies with sub-arcsecond (up to 0″.4) resolution. Stars fainter than 24.5 mag are detected in the $B$ band with an accuracy of ~0.2 mag at an exposure time of 20 min. A distant galaxy of 24.3 ± 0.2 mag has been detected in the SDSS $i$-band in 1 h of exposure time. Optical detection of a radio source observed with GMRT and optical observations of star clusters, AGNs, etc. have been successfully carried out with the 3.6 m DOT during the last three years.

Since the atmospheric seeing varies as $\lambda^{-0.2}$, the average seeing (FWHM) of ~0″.7 observed in the K-band corresponds to ~0″.9 in the visual bands. This agrees with the FWHM values derived from optical images taken with the CCD IMAGER. Observing sub-arcsecond atmospheric seeing at Devasthal even after completion of the telescope indicates that the surrounding infrastructure buildings have not deteriorated natural seeing conditions observed at the site about two decades ago during 1997–1999. The care taken in the design and structure of telescope building has therefore paid a rich dividend in keeping thermal mass of the building very low. The sky performance of Devasthal at NIR wavelengths, known only recently with the TIRCAM2 observations, is encouraging.

This article provides information on about four decades of the long journey from inception to accomplishment of India's largest 3.6 m DOT project. The execution of main project activities like detailed site characterization, land acquisition, necessary approvals from competent authorities and installation of the telescope at Devasthal site was completed during the latter two decades (1996–2016). A total of 29 meetings of the PMB were convened to monitor the progress of various activities, and to find solutions to problems related to the 3.6 m DOT project. The project has been completed only at a few percentage of cost overrun and about two years of time overrun. Due to broad vision and distant view of the Uttarakhand State Government, 50-year-old State Observatory was transferred to DST, GoI. The 3.6 m DOT project, located in Uttarakhand State at Devasthal, Nainital, was completed with fundings from DST, GoI plus 2 million Euros from Belgium which fulfils one of the main objectives of Uttarakhand as the well as the Cabinet decision of GoI taken in 2004.

In order to utilize full astronomical potential of the 3.6 m DOT, it is essential to build modern and complex back-end focal-plane instruments. Due to their great technological complexities, such instruments are generally designed and developed using expertise available across institutions located anywhere in the globe. Development of two such instruments, namely Devasthal Optical Telescope Integral Field Spectrograph (DOTIFS) and TIFR-ARIES Near Infrared Spectrometer (TANSPEC) for the telescope are in the advanced stage. DOTIFS is a multi-object integral field spectrograph. It is being developed, designed and fabricated under the leadership provided by IUCAA. Technical details along with the present status of DOTIFS can be found elsewhere[48–51]. According to design, average throughput of DOTIFS mounted on the 3.6 m DOT is expected to be over 25%. TANSPEC, an optical–NIR medium-resolution spectrograph, has been jointly developed by TIFR and ARIES as part of inter-institutional collaboration. Its wavelength coverage is from 0.55 μm in the optical up to 2.54 μm in NIR. The present status and technical details of this instrument are given elsewhere[39].

In light of the above, it can be said that the 3.6 m DOT observing facility has the potential of providing not only internationally competitive but scientifically valuable optical and NIR observations for a number of frontline galactic and extra-galactic astrophysical research problems, including optical follow-up of GMRT and AstroSat sources. Geographical location of the Devasthal observatory also has global importance for the time domain and multi-wavelength astrophysical studies.

ACKNOWLEDGEMENTS. We thank the reviewer for constructive comments which helped improve the manuscript. We also thank Dr Wahab Uddin and Dr A. K. Pandey for their encouragement and guidance during installation and commissioning of the telescope; staff of ARIES and Devasthal Observatory for their assistance during installation and sky-performance test; Prof. G. Srinivasan for help during telescope building and Prof. Pramesh Rao for valuable contribution in the development of dome control system. We would like to put on record the encouragement and initial funding of Rs 6 lakhs provided by Prof. R. Cowsik (former Director, IIA, Bengaluru) which revived the telescope project after a gap of about six years. The overall guidance and support of the PMB members – Prof. A. S. Kirankumar; Dr T. G. K. Murthy; Prof. S. N. Tandon; Prof. S. Ananthakrishnan; Mr S. C. Tapde; Prof. T. P. Prabhu; Prof. Pramesh Rao and Prof. R. Srinivasan under the Chairmanship of Prof. P. C. Agrawal and Vice-Chairmanship of Prof. G. Srinivasan are acknowledged. Encouragement and guidance provided by Prof. K. Kasturirangan, Prof. S. K. Joshi, Prof. Govind Swarup, late Dr. S. D. Sinvhal, Prof. V. S. Ramamurthy, Prof. T. Ramasami and Prof. J. V. Narlikar in establishing the Devasthal observatory are acknowledged. Approvals, guidance and encouragements rendered by all the members of the GC of ARIES are also acknowledged. The participation of professional companies, viz. AMOS, Belgium; LZOS, Russia; HHV Bengaluru; IMT Govindgarh; Pedvak Hyderabad; PPS Pune, and Schott Germany in the execution of the 3.6 m DOT project is acknowledged. The financial support of 2 million Euros provided by the Belgian Government to the telescope project is appreciated. R.S. thanks the National Academy of Sciences, India (NASI), Allahabad, for award of a NASI Senior Scientist Platinum Jubilee Fellowship; the Alexander von Humboldt Foundation, Germany for the award of Group linkage long-term research program between IIA, Bengaluru and European Southern Observatory, Munich, Germany, and the Director, IIA for hosting.

Received 31 December 2018; revised accepted 31 May 2019

doi: 10.18520/cs/v117/i3/365-381